\documentclass{aastex}
\usepackage{graphicx}

\newcommand{\emaila}{ E-mail: mfullana@mat.upv.es}
\newcommand{\emailb}{ E-mail: aalfonsofaus@yahoo.es}

\begin{document}

\title{Seeable universe and its accelerated expansion: an observational test}

\shorttitle{Seeable universe}
\shortauthors{A. Alfonso-Faus and M. J. Fullana i Alfonso}

\author{Antonio Alfonso-Faus\altaffilmark{1}}
\affil{Escuela de Ingenier\'{\i}a Aeron\'autica y del Espacio, 
Plaza del Cardenal Cisneros, 3,
Madrid, 28040, Spain.\\
\emailb}
\and
\author{M\`arius Josep Fullana i Alfonso\altaffilmark{2}}
\affil{Institut de Matem\`atica Multidisciplin\`aria, 
Universitat Polit\`ecnica de Val\`encia, 
Cam\'{\i} de Vera, 
Val\`encia, 46022, Spain.\\
\emaila}

\begin{abstract}
From the equivalence principle, one gets the strength of the gravitational effect of a mass $M$ on the metric at position r from it. It is proportional to the dimensionless parameter $\beta^2 = 2GM/rc^2$, which normally is $<< 1$. Here $G$ is the gravitational constant, $M$ the mass of the gravitating body, 
$r$ the position of the metric from the gravitating body and $c$ the speed of light. The seeable universe is the sphere, with center at the observer, having a size such that it shall contain all light emitted within it. For this to occur one can impose that the gravitational effect on the velocity of light at $r$ is zero for the radial component, and non zero for the tangential one. Light is then trapped. The condition is given by the equality $R_g = 2GM/c^2$, where $R_g$ represents the radius of the {\it seeable} universe. It is the gravitational radius of the mass $M$. The result has been presented elsewhere as the condition for the universe to be treated as a black hole. According to present observations, for the case of our universe taken as flat ($k = 0$), and the equation of state as $p = - \rho c^2$, we prove here from the Einstein's cosmological equations that the universe is expanding in an accelerated way as $t^2$, a constant acceleration as has been observed. This implies that the gravitational radius of the universe (at the event horizon) expands as $t^2$. Taking $c$ as constant, observing the galaxies deep in space this means deep in time as $ct$, linear. Then, far away galaxies from the observer that we see today will disappear in time as they get out of the distance ct that is $< R_g$. The accelerated expanding vacuum will drag them out of sight. This may be a valid test for the present ideas in cosmology. Previous calculations are here halved by our results.
\end{abstract}


\section{Introduction}

 In \citeyear{b1} \citeauthor{b1} asked himself {\it How large must a universe be in order that it shall contain light emitted within it?} If we change the wording {\it how large} by {\it how strong the gravity field must be} and the word {\it universe} by the word {\it object}, then we have the history of {\it black holes} as follows: first known consideration of light not being able to escape from such an object (not a universe) was given by \citeauthor{b2} in \citeyear{b2}. Eleven years later \cite{b3} addressed the same question. After the publication of Einstein's general relativity \citeauthor{b4} published in \citeyear{b4} a solution to general relativity advancing his well known radius (later referred as the gravitational radius of a mass $M$, $R_g = 2GM/c^2$). About 50 years later it was recognized that black holes were predicted by general relativity. Was \citeauthor{b5} the known author of the name {\it black hole} in \citeyear{b5}, 
referring to the {\it continuous gravitational collapse} of an over-compact mass. It is evident that the first author to treat our universe as a black hole, five years before this name was used, 
was \cite{b1}. It is not necessary to think of a black hole as an over-compact mass. One can have a black hole with any mass, as long as it also has the right size. The biggest mass that one can consider is the mass of the {\it seeable} universe: a universe that contains all light emitted within it. Usually the word {\it visible} universe refers to the visible light from the surface of last scattering, from the recombination epoch (like the CMBR). The word {\it observable} universe refers to the universe we could observe with light or some other means from the initial big-bang. As we will see, here the word {\it seeable} refers to the universe considered as a black hole, like the surface of the event horizon, as noted by \cite{b6}. Now, an average density of the observable universe is of the order of $10^{-29} \ g/cm^3$, about the critical density. The age of the universe t is now believed to be about $1.37 \times 10^{10} \ years$
(\citeauthor{Ben} \citeyear{Ben}). Then multiplying this by the speed of light we get the number $\sim  10^{28} cm$. If we look deep in space we are looking back in time, much  more back the deeper we look. We are limited in going back in time by the amount of $1.37 \times 10^{10} \ years$, as we will see. We are then limited to see deeper than about  $10^{28} cm$. We may roughly calculate a mass for the observable universe by computing the mass from the density inside a sphere of size $\sim  10^{28} cm$. This gives a mass of about $2 \times 10^{56} g$ using the general relativity coefficient for volume $(2 \pi^2)$. Using this value for the mass one gets the value of the gravitational radius of the universe as

\begin{equation}
R_g = 2GM/c^2 \approx 3 \times 10^{28} cm
\label{e1}
\end{equation}

We have called this gravitational radius as the radius of the sphere of the {\it seeable} universe. It has a deep physical meaning. This wording comes from the calculation of the size of a universe that has all its light trapped, as we will see (identified here with the gravitational radius $R_g$). It is then a maximum that cannot be observed in the sense that any light getting close to this limit just turns around and goes back to its origin. It is like being able to see your own back looking deep in space.  But the observational size of the universe is less than that today. When calculating it as the Hubble radius, $c/H$, a conventional way to define an observational universe, it is smaller than the {\it seeable} size $R_g$. We will prove that the radius of the seeable universe increases with time as $R_g \sim t^2$, while the size of observable universe increases as ct (today about $10^{28} cm$). Taking the speed of light as constant this means a linear relation with time. The speed of expansion for the gravitational radius is $(R_g)' \sim t$ and its acceleration is therefore constant, an accelerated universe as has been observed for the last ten years --Type Ia supernova, \cite{b8,b7,b9,b11,b10}; and CMB anisotropies, \cite{b13,b14,b12}--.  Clearly the gravitational radius is larger than the observable radius (about a factor of $2.30$ today) and its expansion going faster drags the far away galaxies out of sight. Looking and identifying a far away galaxy, and looking again after some time long enough, no galaxy would be there: it just disappears. This is a proposed test that may be technically possible in the near future. With the expansion due to the dark energy background of the universe going as $\sim  t^2$, it may be that this happens earlier than thought (\citeauthor{b15} \citeyear{b15}, \citeauthor{b16} \citeyear{b16}). 

\section{Evidence in support of the idea of the seeable universe to be considered as a black hole}

\subsection{Entropy upper bounds}

In \citeyear{b17} \citeauthor{b17}  found an upper bound for the ratio of the entropy $S_B$ to the energy $E = Mc^2$ of any bounded system with effective size $R$:
\begin{equation}
S_B / E  < 2 \pi k R \hbar c
\label{e2}
\end{equation}

About ten years later (\citeauthor{b18} \citeyear{b18}, \citeauthor{b19} \citeyear{b19}) a holographic principle was proposed giving a bound for the entropy $S_h$ of a bounded system of effective size $R$ as

\begin{equation}
S_h  \leq  \pi k c^3 R^2 / \hbar G
\label{e3}
\end{equation}

If the two bounds turn out to be identical $M$ is proportional to $R$ and the system strictly obeys the Schwarzschild condition for a black hole. Also, for the case of a black hole we find that the Hawking and Unruh temperatures are the same. 

\subsection{Consequences of the identification of the two bounds}

Identifying the bounds of equations (\ref{e2}) and (\ref{e3})
\begin{equation}
E 2 \pi k (R /  \hbar c)  =  \pi k (c^3 R^2 / \hbar G)
\label{e4b}
\end{equation}
\noindent
simplifying and using $E=m c^2$, we get
\begin{equation}
2 G M  = c^2 R 
\label{e4}
\end{equation}
This is the condition for the system $(M, R)$ to be a black hole. Then, its entropy is given by the \cite{b20} relation

\begin{equation}
S_H = 4 \pi k / \hbar c \ GM^2
\label{e5}
\end{equation}
that coincides with the two bounds of equations (\ref{e2}) and (\ref{e3}).     
The mass of the universe $M_u$ is a maximum. And so is its size $R$. A bounded system implies a finite value for both. Using present values for 
$Mu \approx 10^{56} g$ and $R \approx 10^{28} cm$ they very closely fulfill the Schwarzschild condition (\ref{e4}). This has been taken as an evidence for the Universe to be considered as a black hole (\citeauthor{b22} \citeyear{b22}, \citeauthor{b23} \citeyear{b23}).

\subsection{The case for the Unruh and Hawking temperatures}
The fact that a black hole has a temperature, and therefore entropy (established by \citeauthor{b20} \citeyear{b20}), implies that an observer at its surface, or event horizon, sees a perfect blackbody radiation, a thermal radiation with temperature $T_H$ given by
\begin{equation}
T_H =  \hbar c^3 / (8 \pi  GM k)
\label{e6}
\end{equation}
where $\hbar$ is the Planck's constant, $c$ the speed of light, $G$ the gravitational constant, $k$ Boltzmann's constant and $M$ the mass of the black hole. This observer feels a surface gravitational acceleration $R''$. According to the \cite{b21} effect an accelerated observer also sees a thermal radiation at a temperature $T_U$, proportional to the acceleration $R''$ and given by

\begin{equation}
T_U =  \hbar R" / (2 \pi  c k)
\label{e7}
\end{equation}
 
Based upon the similarity between the mechanical and thermo dynamical properties of both effects, equations (\ref{e6}) and (\ref{e7}), 
we identify both temperatures and find the relation:

\begin{equation}
R" =  c^4 / (4GM)
\label{e8}
\end{equation}

Identifying the Unruh acceleration to the surface gravitational acceleration: 

\begin{equation}
R"  =  GM / R^2
\label{e9}
\end{equation}
and substituting in equation (\ref{e8})  we finally get
\begin{equation}
2GM / c^2= R
\label{e10}
\end{equation}

This is the condition for a black hole. Since the Hawking temperature refers to a black hole this result confirms the validity of the identification of the two temperatures, equations (\ref{e6}) and (\ref{e7}), as well as the interpretation of the Unruh acceleration in equation (\ref{e9}).

\section{The gravitational radius of the universe that traps all light}

From the equivalence principle, one gets the strength of the gravitational effect of a mass $M$ on the metric at position $r$ from it (\citeauthor{b1}
\citeyear{b1}). It is proportional to the dimensionless parameter $\beta^2 = 2GM/rc^2$, which normally is $<< 1$. Here $G$ is the gravitational constant, $M$ the mass of the gravitating body, $r$ the position of the metric from the gravitating body and $c$ the speed of light. The {\it seeable} universe is the sphere, with center at the observer, having a size such that it shall contain all light emitted within it. For this to occur, one can impose that the gravitational effect on the velocity of light at $r$ give a zero value for the radial component, and non zero for the tangential one  (\citeauthor{b1} \citeyear{b1}). Then, light is always inside the seeable universe, trapped within it. 
The condition is given by the equality $R_g = 2GM/c^2$, where $R_g$ represents the radius of the {\it seeable} universe. 
It is the gravitational radius of the mass $M$. We will prove in the next section that the cosmological equation of state $p = \omega \rho c^2$, 
with the present observed value of $\omega \approx -1$, implies that the gravitational radius $R_g$ varies as $\sim t^2$.  

\section{Einstein/Friedmann cosmological equations and the equation of state $p = \omega \rho c^2$}

The Einstein tensor $G_{\mu \nu}$ has a 4D zero divergence by construction:
\begin{equation}
G^{\mu}_{\mu ; \nu} = 0
\label{e11}
\end{equation}

The tensor $G_{\mu \nu}$ is a geometrical object and is the left hand side of the field equations. Then, the covariant divergence of the second term, i. e., $8 \pi (G/c^4) T^{\mu \nu}$, $T^{\mu \nu}$ the energy-momentum tensor, must be zero. The resultant equation has been found under the full generalization of considering $G$, $c$ and the cosmological $\Lambda$ as time variables 
(\citeauthor{b24} \citeyear{b24}).  We repeat it here:

\begin{equation}
\rho ' / \rho + 3 (\omega + 1) H + \Lambda ' c^4 / 8 \pi G \rho + G' / G - 4 c' / c = 0
\label{e12}
\end{equation}
where $\rho$ is here energy density, $H$ the Hubble parameter $H = a(t)'/a(t)$, $a(t)$ the cosmological scale factor as is defined in the Robertson-Walker metric, and $\omega$ defines the equation of state. Considering the cosmological constant $\Lambda$  as a true constant one can integrate equation
(\ref{e12}) giving,

\begin{equation}
(G \rho / c^4) a(t)^{3(\omega +1)} = \hbox{constant}
\label{e13}
\end{equation}

Using the conversion $(G \rho / c^4) \propto (G / c^2) [M / a(t)^3]$, since $ \rho / c^2 $ is mass density, i.e. $ M / a(t)^3$, 
the result (\ref{e13})  is transformed to

\begin{equation}
(G M / c^2) a(t)^{3\omega} = \hbox{constant}
\label{e14}
\end{equation}
                         
And using the definition of the gravitational radius $R_g$ as the Schwarzschild radius we get from equation (\ref{e14})
\begin{equation}
R_g a(t)^{3\omega} = \hbox{constant}
\label{e15}
\end{equation}

Hence, the gravitational radius, the {\it seeable} size of the universe as a black hole, depends only on the cosmological scale factor $a(t)$ and the equation of state parameter $\omega$. With the present value for the parameter $\omega = -1$ we have the solution from equation (\ref{e15}) as

\begin{equation}
R_g  = a(t)^{3} \ \hbox{constant}
\label{e16}
\end{equation}

We use now the Friedmann equation that, together with equation (\ref{e15}), completely defines the cosmological model (the integration of the two Einstein cosmological equations):

\begin{equation}
(H)^2 = [a(t)' / a(t)]^2 = 8 \pi G \rho / 3 = G M / [a(t)]^3 \times \hbox{constant}
\label{e17}
\end{equation}

Whatever the time dependence of $G$ and $M$ may be, we prove now that their product $GM$ must be a constant. This comes from the consideration that the entropies in equations (\ref{e2}), (\ref{e3}) and (\ref{e5}) must be an extensive property, proportional to mass $M$.

We know that $G$ is always associated with a mass. Then the constancy of mass can be substituted by the constancy of $GM$. 
The Friedman equation (\ref{e17}) is then,

\begin{equation}
a(t) [a(t)']^{2} = \hbox{constant}
\label{e18}
\end{equation}
which is the well known Einstein-de Sitter model with solution:
\begin{equation}
a(t) / a(t_0)'] = (t/t_0)^{2/3}
\label{e19}
\end{equation}

Recently  (\citeauthor{b25} \citeyear{b25})
it has been shown that the probability that the universe emerges right after a {\it big trip}, thus avoiding a {\it big rip}, in a state of initial constant $\omega_0$, will be maximal if and only if one has $\omega_0 = -1/3$. From equation (\ref{e15}) we see that this value corresponds to a gravitational radius for the universe expanding as the cosmological factor $a(t)$ given in equation (\ref{e18}), an initial Einstein-de Sitter state of the universe expanding as in equation (\ref{e19}).  After this we have today strong evidence that the value of $\omega$ is $-1$, at least as far back as redshift $z \approx 1$. Again, going to equation (\ref{e15}) we see that the gravitational radius of the universe expands now as, 

\begin{equation}
R_g(t) / R_g(t_0) = [a(t) / a(t_0)]^{-3 \omega} = [a(t) / a(t_0)]^{3} = (t/t_0)^2
\label{e20}
\end{equation}
 
Then, the gravitational radius of the universe starts with a time dependence as given by the Einstein-de Sitter universe, equation (\ref{e19}) with the equation of state $p = -1/3 \rho c^2$ and evolves in such a way that the dark matter and energy components make the equation very close to 
$p = - \rho c^2$ as of today. The observable size of the universe going as $ct$ implies that all contents in the universe are dragged faster than this, 
$\sim t^2$  as in equation (\ref{e20}).

\section{The proposal for a test of the present cosmological models}

\begin{figure}
\includegraphics[width=8 cm,height=8 cm]{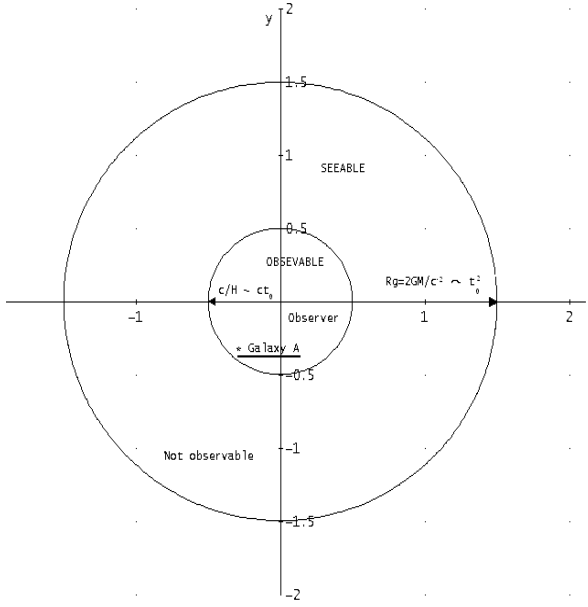}
 \caption{ \label{f1} Position of galaxy A at $t_0$ in the observable universe.}
 \end{figure}

\begin{figure}
\includegraphics[width=8 cm,height=8 cm]{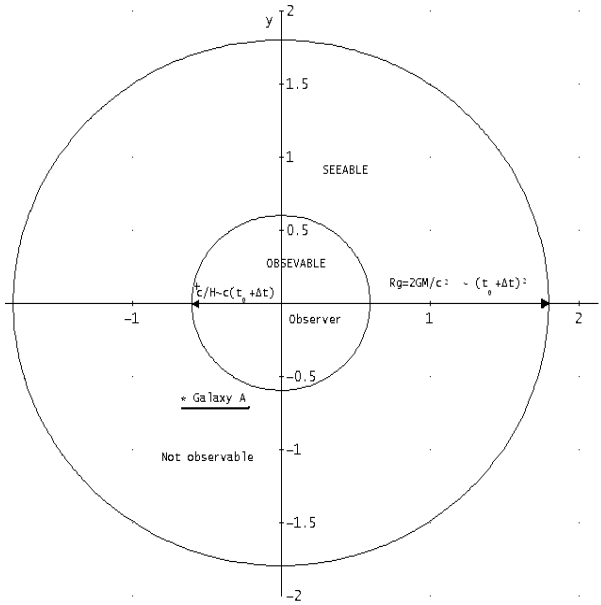}
 \caption{ \label{f2}Position of galaxy A at $t_0 + \Delta t$ out of the observable universe.}
\end{figure}

In our approach galaxies are physically dragged by expansion as $\sim t^2$ which means that they are disappearing when they are dragged beyond the distance $c/H \approx ct$. 
By looking at a particular identifiable galaxy deep in space, we could check its disappearance after some finite time. This implies a repetition of observations of the same galaxy deep in space hoping to find at some later time that the pinpointed galaxy is no longer there. 
Figures \ref{f1} and \ref{f2} present this situation. 
\citeauthor{b15} presented in \citeyear{b15} arguments to expect objects to become out of causal contact.  Later \cite{b16}, as well as \cite{b26}, pointed out this possibility about ten years ago. 
In this direction an out of sight possibility for far away galaxies has been pointed out by \cite{b6}. The calculations made, as \cite{b26} presented, must be halved in our case due to the $\sim t^2$ dependence of the dragging background presented here. We also present it under a different point of view, with the same basic idea of getting out of sight galaxies in time due to accelerated expansion. Our conclusion on the dragging effect going as $\sim t^2$ halves the calculations presented by these authors.

\section{Conclusions}

The universe can be seen as a black hole. From each observer in it we can interpret that he is at the centre of a sphere, with the Hubble radius, that would be the observable universe, but there is the gravitational radius that is larger than the observable one. 
The isotropic acceleration present at each point in the universe, and given by equation (\ref{e9}), implies that there is no distortion for the spherically distributed acceleration if we consider the strict conditions of the cosmological principle. However, the presence of a nearby important mass, like the sun, in any point in interplanetary space will distort this spherically symmetric picture. With respect to the probes Pioneer 10/11, that detected an anomalous extra acceleration towards the sun of value $(8.74 \pm 1.33) \times 10^{-8} cm/s^2$ 
(\citeauthor{b27} \citeyear{b27}), we can see that this value is only a bit higher than the one predicted by equation (\ref{e9}), 
which is about $7.7 \times 10^{-8} cm/s^2$. This difference is an effect that can be explained by the presence of the sun converting the isotropic acceleration to an anisotropic one.
For about ten years we have good evidence that the universe is expanding in an accelerated way. We have found here that the present observations, that favours the value $\omega = -1$ for the equation of state, imply an expanding law of the type $R_g \sim t^2$ giving a constant acceleration. Thus, the gravitational radius Rg, that can be interpreted as the size of the universe as a black hole, trapping all light gives the maximum {\it seeable} size for the universe. The Hubble radius, $c/H$, can be interpreted as the maximum observable size for the universe since it is limited by its age, the distance travelled by light going backwards in time up to the initial stages. The background {\it vacuum} is then expanding at a faster rate than it can be observed. And this {\it vacuum} is the frame for all cosmological structures, like the galaxies. Then, galaxies are physically dragged by expansion as $\sim t^2$ which means that they are disappearing when they are dragged beyond the distance $c/H$. By looking at a particular identifiable galaxy very deep in space, we could check its disappearance after some finite time. The past calculations for this to occur are here halved by the expansion law $ \sim t^2$.

\end{document}